\documentclass[12pt,thmsa]{article}
\usepackage{amsfonts}
\usepackage{graphicx}
\usepackage{epsfig}
\usepackage{graphics}

\makeatletter
\newcommand{\row}[1]{\mathord{\buildrel{\lower3pt\hbox{$\scriptscriptstyle\rightarrow$}}\over #1}}

\newcommand{\dyadic}[1]{\mathord{\dyadic@rrow{#1}}}
\newcommand{\dyadic@rrow}[1]{
\begin{picture}(12,12)(-1,0)
\put(-2,12){\makebox(0,0)[t]{$\scriptscriptstyle\downarrow$}}
\put(-2,12){\makebox(0,0)[l]{$\scriptscriptstyle\longrightarrow$}}
\put(5,0){\makebox(0,0)[b]{$#1$}}
\end{picture}
}
\newcommand{\bra}[1]{\bigl\langle #1 \bigr|}
\newcommand{\ket}[1]{\bigl| #1 \bigr\rangle}

\input{tcilatex}
\begin{document}

\begin{center}

Some properties of multi-qubit systems under  effect of the
Lorentz
transformation\\
N. Metwally
\\

{\large {\footnotesize  Math. Dept. Faculty of Science, Aswan University, Aswan, Egypt }}
\end{center}

\begin{abstract}

Effect of Lorentz transformation on  some properties  of
multi-qubit systems is investigated.  It is shown that, properties
like, the fidelity and entanglement decay as the Wigner's angles
increase, but can be  improved, if all the transformed particles
are  transformed with the same Wigner's angles. The upper bounds
of the average capacity of  the GHZ state increases while it
decreases and more robust with the  W-state as the Wigner's angle
of the observer decreases. Under Lorentz transformation, the
tripartite states  transform  into another equivalent states and
hence no change on the efficiency of these states to perform
quantum information tasks.

\end{abstract}

{\bf keywords:} Lornetz transformation, Mutlti-qubits, Entanglement, Fidelity, Capacity

\section{introduction}
It is well known that entanglement represents the corner stone of
most applications of quantum  information
\cite{Nilson,Ahmed2012,Metwally2013}. There are several studies
 devoted to investigate the possibility of generating
entanglement between different types of particles
\cite{Aty2001,Barry2010,Aty2005}. Quantifying the degree of
entanglement which may be generated between these particles
assimilates another area for many researchers, where several
measurements of entanglement have been found. These measures
depend on the type of the generated entangled state, pure or
mixed, in two dimensions or higher. The most common measures for
two qubit systems are, the concurrence \cite{Wootter89},
entanglement of formation \cite{Benn96,Hill97} and  negativity
\cite{Zyc,Vidal02} etc, while for the  tripartite states, tangle
represents an acceptable measure \cite{Coff2000}.

Recently, it has been shown that entanglement is treated from
relativistic point of view. For example, the releativistic
entanglement of two massive particles is investigated by  N.Friis
et.al \cite{Friis2010}. Saldanh and Vedral \cite{Pablo2012} have
discussed the spin quantum correlations of relativistic particles.
The behavior of the spin fidelity of the three qubit Greenberge
-Horne -Zeingler and W-state under Lorentz transformation is
studied by Esfahani and M. Aghaee \cite {Nasr2011}. The change of
entanglement under the effect of Lorentz transformation of a two
spin-one particle system is studied by Ruiz and N. Achar
\cite{Ruiz2-13}.

In this contribution, we investigate the effect of Lorentz
transformation on two classes of multi-qubit systems: GHZ and
W-states. The behavior of the fidelities and the channel
capacities are investigated for different values of Wigner angles.
Due to the effect of Lorentz transformation the entanglement of
these multi-qubit systems decays. Therefor, we investigate the
robustness of these states under the action of Lorentz
transformation.

The paper is organized as follows: In Sec. 2, the  suggested model
and its  evolution under the effect of the  Lorentz transformation
is discussed. Sec. 3,  is devoted to investigate the immutability
of the multi-qubit states by discussing the  behavior of the
fidelities and  the average capacities   of these states. The
amount of entanglement of the transformed states is quantified by
using the three-tangle \cite{Coff2000} in Sec. 4. Finally, our
results are summarized in Sec. 5.

\section{The Suggested Model}
We assume that,  a source supplies a three users with a three
massive particles. It is assumed that, the spin part is given by
GHZ or W-states, while the momentum part is a superposition
between the three qubits. Therefore, the total  system  can be
described by the following state \cite{Jordan2007}
\begin{equation}\label{ini}
\ket{\psi_{system}}=\ket{\psi_{mom}}\ket{\psi_{spin}},
\end{equation}
where, $\ket{\psi_{mom}}$ represents the state vector of the
momentum given by
\begin{equation}
\ket{\psi_{mom}}=\sin\alpha\ket{p_1^-,p_2^-,p_3^-}+\cos\alpha\ket{p_1^+,p_2^+,p_3^+},
\end{equation}
while the spin part is given by one of the for following states
\begin{eqnarray}\label{iniS}
 \ket{\psi_{spin}}&=& \left\{ \begin{array}{ll}
\ket{\psi_{g}}=\frac{1}{\sqrt{2}}(\ket{000}+\ket{111}),&  \\
\ket{\psi_{g'}}=\frac{1}{\sqrt{2}}(\ket{000}-\ket{111}),&  \\
\ket{\psi_{w}}=\frac{1}{\sqrt{3}}(\ket{100}+\ket{010}+\ket{001}),& \\
\ket{\psi_{w'}}=\frac{1}{\sqrt{3}}(\ket{110}+\ket{101}+\ket{011}),& \\
\end{array} \right.
\end{eqnarray}
The state vector $\ket{\psi_{g,g'}}$  and  $\ket{\psi_{w,w'}}$ are
known by Greenberge-Horn-Zeilinger (GHZ) and W-states
respectively. The computational basis $"0$ and $"1"$  represent
spins polarized up and down along the z-axis. The action of an
arbitrary Lorentz transformation $\Lambda$ on the initial state
$\ket{\psi_{system}}$ is given by \cite{Friis2010,Jordan2007}
\begin{equation}
\Lambda\ket{\psi_{system}}=
\sum_{i}^{3}D(W(\Lambda,p_i))\ket{\psi_{mom}}\ket{\psi_{spin}}
\end{equation}
where, $D(W(\Lambda,p_i))$ represents the Wigner rotation operator
is defined by
\begin{equation}\label{uni}
D(W(\Lambda,p_i))=\cos\frac{\Omega_i}{2}+i\row{\sigma}\cdot\row{n}\sin\frac{\Omega_i}{2}
\end{equation}
where $\row\sigma=(\sigma_x,\sigma_y,\sigma_z)$, and  $\Omega_i,
i=1,2,3$ are the Wigner angles. The operator
$W(\lambda,p)=L^{-1}(\Lambda p)\Lambda L(p)$ is the Winger's
little group element, $L(p)$  is the standard boost that transform
a particle of mass $m$ from the rest to four momenta $\row{p}$
\cite{Friis2010,Nasr2011}. If the momenta  are chosen such that
$\row{p}_1=\row{p_2}=\row{p_3}=p\row{e}_z$, i.e, the momenta are
polarized in the $z-$axis, then in  the computational basis $"0"$
and $"1"$, the unitary operator(\ref{uni}) takes the following
form,
\begin{eqnarray}\label{Lornetz}
D(W(\Lambda,p_i))&=&\mathcal{C}_i(\ket{0}\bra{0}+\ket{1}\bra{1})+\mathcal{S}_i(\ket{1}\bra{0}-\ket{0}\bra{1})
\end{eqnarray}
where $\mathcal{C}_i=\cos\frac{\Omega_i}{2},
\mathcal{S}_i=\sin\frac{\Omega_i}{2}, i=1,2,3$. Using the initial
state (\ref{ini}), the Lorentz transformation (\ref{Lornetz}) and
tracing out the momentum degree of freedom, one gets the evolution
of the spin part.

With the source supplies the user with a GHZ state,
 of the type $\ket{\psi_g}$ as defined in (\ref{iniS}). It has been shown that this class of
entangled  states  turns into separable  states if one of its
particle is traced out \cite{Dur}. Under the effect of the Lorentz
transformation (\ref{Lornetz}) the  initial state $\ket{\psi_g}$
is transformed into,
\begin{equation}\label{GHZF}
\ket{\psi_{g_f}}=\mathcal{A}_1\ket{000}+\mathcal{A}_2\ket{001}+\mathcal{A}_3\ket{010}+\mathcal{A}_4\ket{011}+
\mathcal{A}_5\ket{100}+\mathcal{A}_6\ket{101}+\mathcal{A}_7\ket{110}+\mathcal{A}_8\ket{111},
\end{equation}
where
\begin{eqnarray}
\mathcal{A}_1&=&\frac{1}{\sqrt{2}}(\mathcal{C}_1\mathcal{C}_2\mathcal{C}_3-\mathcal{S}_1\mathcal{S}_2\mathcal{S}_3),\quad
\mathcal{A}_2=\frac{1}{\sqrt{2}}(\mathcal{C}_1\mathcal{C}_2\mathcal{S}_3+\mathcal{S}_1\mathcal{S}_2\mathcal{C}_3), \nonumber\\
\mathcal{A}_3&=&\frac{1}{\sqrt{2}}(\mathcal{C}_1\mathcal{S}_2\mathcal{C}_3+\mathcal{S}_1\mathcal{C}_2\mathcal{S}_3),\quad
\mathcal{A}_4=\frac{1}{\sqrt{2}}(\mathcal{C}_1\mathcal{S}_2\mathcal{S}_3-\mathcal{S}_1\mathcal{C}_2\mathcal{C}_3),\nonumber\\
\mathcal{A}_5&=&\frac{1}{\sqrt{2}}(\mathcal{S}_1\mathcal{C}_2\mathcal{S}_3+\mathcal{C}_1\mathcal{S}_2\mathcal{CS}_3),\quad
\mathcal{A}_6=\frac{1}{\sqrt{2}}(\mathcal{S}_1\mathcal{C}_2\mathcal{S}_3-\mathcal{C}_1\mathcal{S}_2\mathcal{C}_3),\nonumber\\
\mathcal{A}_7&=&\frac{1}{\sqrt{2}}(\mathcal{S}_1\mathcal{S}_2\mathcal{C}_3-\mathcal{C}_1\mathcal{C}_2\mathcal{S}_3),\quad
\mathcal{A}_8=\frac{1}{\sqrt{2}}(\mathcal{C}_1\mathcal{C}_2\mathcal{C}_3+\mathcal{S}_1\mathcal{S}_2\mathcal{S}_3).
\end{eqnarray}

On the other hand, if we assume that the  source supplies  the
three users with W-state, then under the effect of Lorentz
transformation, the initial  $\ket{\psi_w}$ state turns into the
state $\ket{\psi_{w_f}}$,
\begin{equation}\label{GHZF}
\ket{\psi_{w_f}}=\mathcal{B}_1\ket{000}+\mathcal{B}_2\ket{001}+\mathcal{B}_3\ket{010}+\mathcal{B}_4\ket{011}+
\mathcal{B}_5\ket{100}+\mathcal{B}_6\ket{101}+\mathcal{B}_7\ket{110}+\mathcal{B}_8\ket{111}
\end{equation}
where,
\begin{eqnarray}
\mathcal{B}_1&=&\frac{1}{\sqrt{3}}\big(\mathcal{S}_1\mathcal{C}_2\mathcal{S}_3+\mathcal{S}_1\mathcal{S}_2\mathcal{C}_3
+\mathcal{C}_1\mathcal{S}_2\mathcal{S}_3\bigl),\quad
\mathcal{B}_2=\frac{1}{\sqrt{3}}\big(\mathcal{S}_1\mathcal{S}_2\mathcal{S}_3-\mathcal{S}_1\mathcal{C}_2\mathcal{C}_3
-\mathcal{S}_1\mathcal{C}_2\mathcal{C}_3\bigl)
 \nonumber\\
\mathcal{B}_3&=&\frac{1}{\sqrt{3}}\big(\mathcal{S}_1\mathcal{S}_2\mathcal{S}_3-\mathcal{S}_1\mathcal{C}_2\mathcal{C}_3
-\mathcal{S}_1\mathcal{C}_2\mathcal{C}_3\bigl),\quad
\mathcal{B}_4=\frac{1}{\sqrt{3}}\big(\mathcal{C}_1\mathcal{C}_2\mathcal{C}_3-\mathcal{S}_1\mathcal{S}_2\mathcal{C}_3
-\mathcal{S}_1\mathcal{C}_2\mathcal{S}_3\bigl)
 \nonumber\\
\mathcal{B}_5&=&\frac{1}{\sqrt{3}}\big(\mathcal{S}_1\mathcal{S}_2\mathcal{S}_3-\mathcal{C}_1\mathcal{S}_2\mathcal{C}_3
-\mathcal{C}_1\mathcal{C}_2\mathcal{C}_3\bigl),\quad
\mathcal{B}_6=\frac{1}{\sqrt{3}}\big(\mathcal{C}_1\mathcal{C}_2\mathcal{C}_3-\mathcal{S}_1\mathcal{S}_2\mathcal{C}_3
-\mathcal{C}_1\mathcal{S}_2\mathcal{S}_3\bigl)
 \nonumber\\
 \mathcal{B}_7&=&\frac{1}{\sqrt{3}}\big(\mathcal{C}_1\mathcal{C}_2\mathcal{C}_3-\mathcal{S}_1\mathcal{C}_2\mathcal{S}_3
-\mathcal{C}_1\mathcal{S}_2\mathcal{S}_3\bigl),\quad
\mathcal{B}_8=\frac{1}{\sqrt{3}}\big(\mathcal{S}_1\mathcal{C}_2\mathcal{C}_3+\mathcal{C}_1\mathcal{S}_2\mathcal{C}_3
+\mathcal{C}_1\mathcal{C}_2\mathcal{S}_3\bigl).
 \nonumber\\
\end{eqnarray}
On the other hand, if we consider that  the spin part is given by
$\ket{\psi_{g'}}$ or $\ket{\psi_{w'}}$, one gets a similar
expression for the final states of $\ket{\psi_{g_f}}$ and
$\ket{\psi_{w_f}}$ which are given by Eq.(7) and Eq.(9)
respectively.

In the following sections, we investigate the effect of Lorentz
transformation on some  properties related to the above model
within  the context of quantum information and
computation.Specifically, we examine the three quantities (i)
fidelity, which measures the closeness of the initial and final
states, (ii) the channel capacity, which measures how much
information can be sent by using these final state, and the (iii)
entanglement, which quantify the degree of correlation between the
subsystems of these states.

\section{Robustness of the transformed states}

\subsection{Fidelity}
 Now, it is important to shed the light on the  robustness of
the these multi-qubit states against the  Lorentz transformation.
One of the important properties of the GHZ state is investigating
the behavior of its fidelity. The closeness of   of the initial
GHZ state $\rho_{g}=\ket{\psi_g}\bra{\psi_g}$ to the final state
$\rho_{g_f}=\ket{\psi_{g_f}}\bra{\psi_{g_f}}$ is defined by
\begin{equation}
\mathcal{F}_{{g}^{\pm}}=\frac{1}{\sqrt{2}}(|\mathcal{A}_1|^2\pm
\mathcal{A}_1\mathcal{A}_8^*
\pm\mathcal{A}_8\mathcal{A}_1^*+|\mathcal{A}_8|^2),
\end{equation}
where $\mathcal{F}_{g^+}$,   and $\mathcal{F}_{g^-}$ represent the
fidelity of $\ket{\psi_{g_f}}$ with respect to $\ket{\psi_g}$
$\ket{\psi_g'}$ respectively.

The behavior of the fidelity $\mathcal{F}_{g^{\pm}}$, is described
in Fig. $1a$, where it is  assumed that the Wigner angles
$\Omega_1=\Omega_2=\Omega'$. It is clear that, at
$\Omega'=\Omega_3=0$ the fidelity $\mathcal{F}_{g^+}=1$ (maximum).
However as the Lorentz transformation is acted on,  the fidelity
decreases and completely vanishes at $\Omega'=0$ and
$\Omega_3=\pi$. The vanishing of the fidelity  is also  seen at
$\Omega'=\pi$, $\Omega_3=0$ and at $\Omega'=\Omega_3=\pi$. On the
other hand, the fidelity $\mathcal{F}_{g^{\pm}}$ of the  GHZ state
is maximum i.e., $\mathcal{F}_{g^+}=1$ when $\Omega'=2\pi$,
$\Omega_3=0$ and at $\Omega'=\Omega_3=2\pi$.

If we consider the GHZ state  given by
$\ket{\psi_g^-}=\frac{1}{\sqrt{2}}(\ket{000}-\ket{111})$, then the
fidelity $\mathcal{F}_{g^-}=tr\{\rho_{g_f}\rho_{g^-}\}$. The
behavior of this fidelity is displayed in Fig.(1b). It is clear
that, $\mathcal{F}_{g^-}=0$ is at $\Omega'_1=\Omega_3=0$ or
$2\pi$. The maximum value of  $\mathcal{F}_{g^-}$ is reached at
$\Omega'=\Omega_3=\pi$.

Fig. 1c, displays the behavior of the fidelity of the GHZ state
when exposed to Lorentz transformation, where we assume that
$\Omega_1=\Omega_1=\Omega_3=\Omega'\in[0,2\pi]$. It is clear that
at $\Omega'=0$, the fidelity $\mathcal{F}_g$ is maximum. As
$\Omega'$ increases the fidelity decreases to reach its minimum
value at $\Omega'=\pi$. However as $\Omega'$ increases the
fidelity increases gradually to reach its maximum value at
$\Omega'=2\pi$.
\begin{figure}[t!]
  \begin{center}
  \includegraphics[width=18pc,height=14pc]{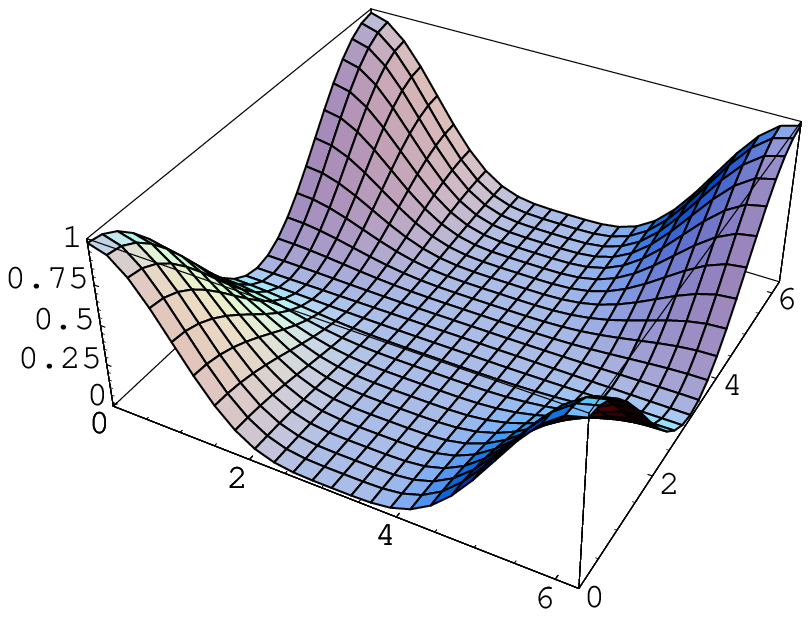}~\quad
   \includegraphics[width=18pc,height=14pc]{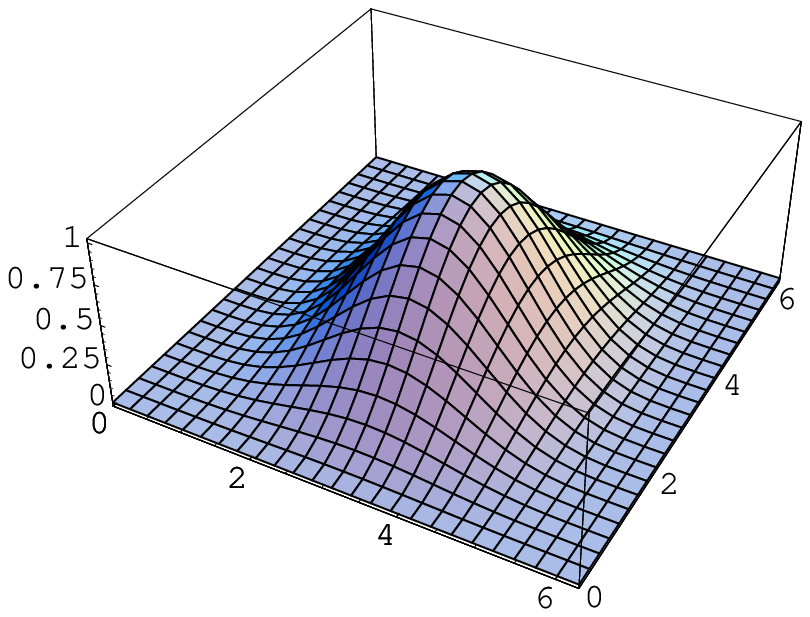}
   \put(-460,80){$\mathcal{F}_{g^+}$}
  \put(-390,20){$\Omega'$}
 \put(-255,50){$\Omega_3$}
  \put(-228,75){$\mathcal{F}_{g^-}$}
  \put(-160,20){$\Omega'$}
 \put(-20,40){$\Omega_3$}\\
  \includegraphics[width=15pc,height=10pc]{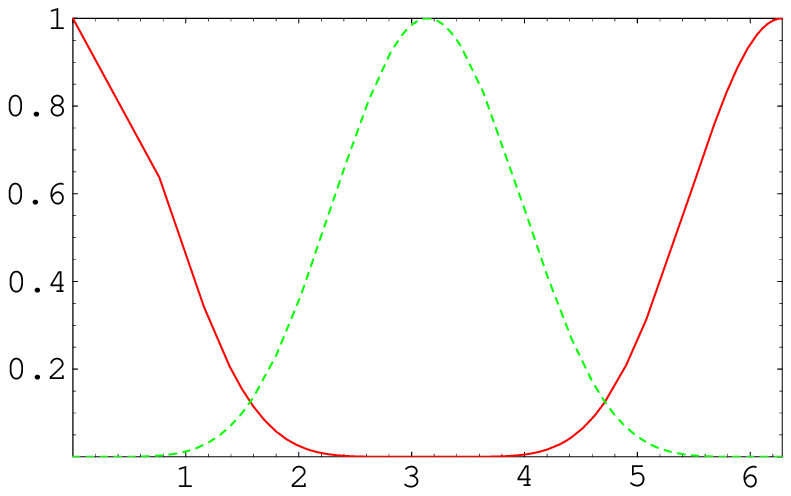}
      \put(-90,-10){$\Omega'$}
 \put(-210,60){$\mathcal{F}_{g^{+},g^-}$}
     \caption{The fidelity of the GHZ states where $\Omega_1=\Omega_2=\Omega_3=\Omega$.
     The solid and dot curves for $\mathcal{F}_{g^+}$ and
     $\mathcal{F}_{g^-}$, respectively. }
       \end{center}
\end{figure}

\begin{figure}
  \begin{center}
  \includegraphics[width=18pc,height=14pc]{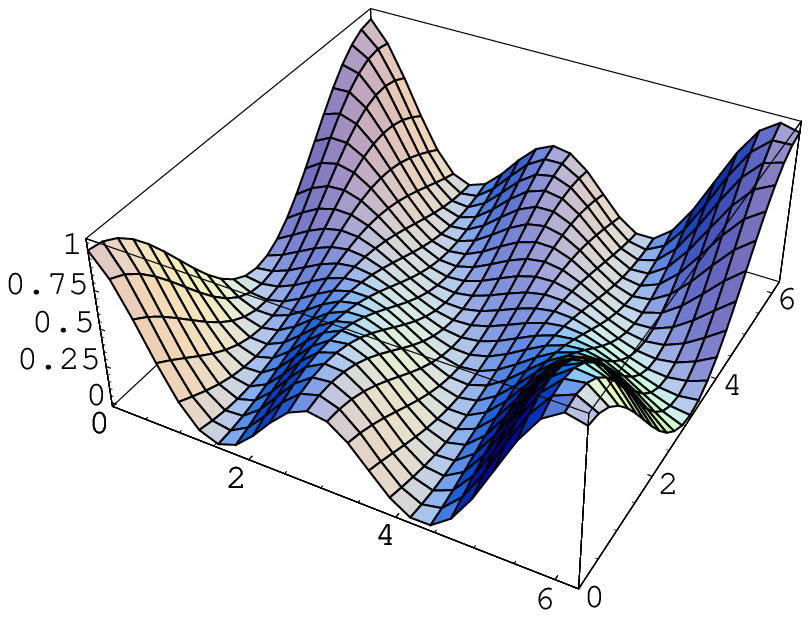}~\quad
     \includegraphics[width=18pc,height=14pc]{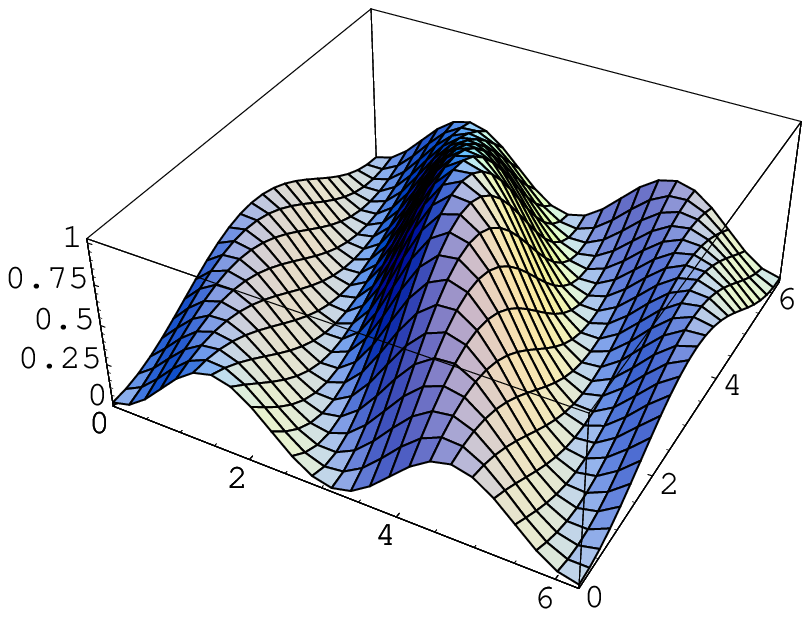}
  \put(-460,80){$\mathcal{F}_w$}
  \put(-390,20){$\Omega'$}
 \put(-255,50){$\Omega_3$}
  \put(-228,75){$\mathcal{F}_w'$}
  \put(-160,20){$\Omega'$}
 \put(-20,40){$\Omega_3$}\\\
  \includegraphics[width=15pc,height=10pc]{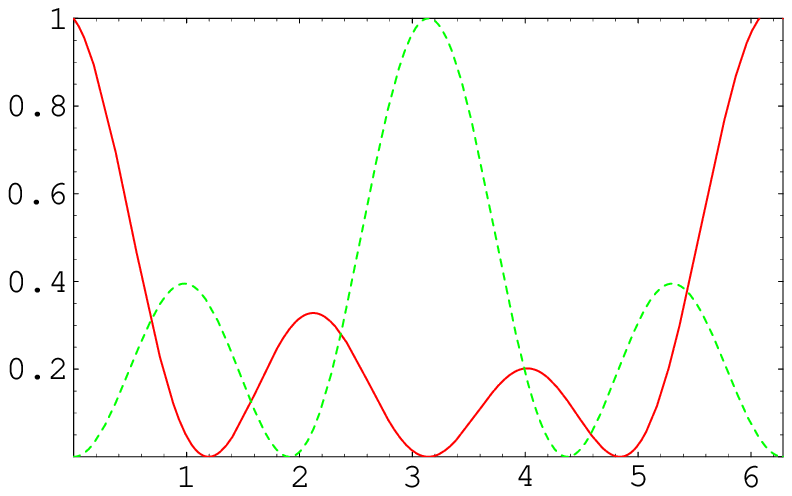}
     \put(-90,-10){$\Omega'$}
 \put(-210,60){$\mathcal{F}_{w,w'}$}
     \caption{The fidelity for the  W-states, where $\Omega_1=\Omega_2=\Omega_3=\Omega'$. The solid and dot curves for $\mathcal{F}_w$ and $\mathcal{F}_{w'}$ respectively.  }
       \end{center}
\end{figure}

For W-state the behavior of the fidelity is described in Fig.$2$,
where we assume that, the particles are transformed with the same
Wigner angles for GHZ state. It is clear that, at
$\Omega'=\Omega_3=0$ i.e., before switching on the Lorentz
transformation, the fidelity $\mathcal{F}_w=1$ (maximum). However
as $\Omega'$ or $\Omega_3$ increase the fidelity decreases
smoothly to vanish completely for the first time  at
$\Omega'\simeq \frac{3\pi}{8}$. The fidelity, $\mathcal{F}_w$
re-increases again to reach its upper bound at
$\Omega'=\frac{2\pi}{3}$. On the other hand, as $\Omega_3$
increases the fidelity $\mathcal{F}_w$ vanishes  once for
$\Omega_3\in[0,2\pi]$. It is clear that the fidelity of W-state
reaches its maximum values either at $\Omega'=\Omega_3=0,2 \pi$ or
$\Omega'=0$ and $\Omega_3=2\pi$ or $\Omega'=\pi$ and $\Omega_3=0$.

It is well know that the  W-state has another form defined by
$\ket{\psi_w'}=\frac{1}{\sqrt{3}}(\ket{011}+\ket{101}+\ket{110})$.
In Fig.(2b), we plot the fidelity $\mathcal{F}_{w'}$ of the state
(8) with respect to the $\ket{\psi_w'}$.  The general behavior of
the fidelity  $\mathcal{F}_{w'}$ shows that as
$\mathcal{F}_{w}=1$( as shown in Fig.(2a)), the fidelity
$\mathcal{F}_{w'}=0$ (minimum). This behavior is depicted for all
values of $\Omega'$ and $\Omega_3$

Fig.2c,  summarizes  clearly these results which depicted in Figs.
$(2a\& 2b)$. It is clear that at   $\Omega'=0$ the fidelity
$\mathcal{F}_w=1$(maximum),   $\mathcal{F}_w'=0$ (minimum).
However,  as $\Omega'$ increases the fidelity $\mathcal{F}_w$
completely vanishes twice  at $\Omega'=\frac{3\pi}{8}$ and
$\frac{2\pi}{3}$, while $\mathcal{F}_w'$ reaches its maximum
bounds ($\mathcal{F}_w'=1$). This behavior  completely changes
i.e., at the values of $\Omega'$ which maximize  $\mathcal{F}_w$,
the fidelity $\mathcal{F}_w'$ is minimized.

From Figs.(1) and (2), one concludes that, although  the Lorentz
transformation causes a decay of the fidelity, it generates an
equivalent state which has a maximum fidelity when the initial one
has a minimum fidelity. Therefore, the Lorentz transformation has
no effect on the efficiency of these classes of the tripartite
states within the context of quantum information.

\subsection{The average capacity} In this subsection, we investigate
the  ability of using the tripartite state under Lorentz
transformation to  send information. For this aim we quantify the
average  capacity of the final transformed states. For the state
$\rho_{abc}$, we have three possible channels between  each two
users:$\rho_{ab}, \rho_{ac}$ and $\rho_{bc}$, where $a,b$ and $c$
represent the three users. This means that each two users share a
two-qubit state. The channel capacity of a two qubit state is
defined as,

\begin{equation}
\mathcal{C}^{(k)}_p(\rho_{ij})=log_i
D+\mathcal{S}(\rho_i^{(k)})-\mathcal{S}(\rho_{ij}^{(k)}),
\end{equation}
where  $\rho_i=tr_j\{\rho_{ij}\}$, $D=2$ is the dimension of
$\rho_i$, $i=a,b,c,  ij=ab,ac,bc$
 and $\mathcal{S}(.)$ is the von Numann entropy. The superscript $k$ stands for the GHZ  or W- state. For tripartite
state the average capacity between the
 three users can be considered  as a measure of the  average capacity of the state $\rho_{abc}$
 is defined as,
 \begin{equation}
\bar{\mathcal{C}_p}^{(k)}(\rho_{abc})=\frac{1}{3}\Bigl(\mathcal{C}^{(k)}_p(\rho_{ab})+\mathcal{C}^{(k)}_p(\rho_{bc})
+\mathcal{C}^{(k)}_p(\rho_{ac})\Bigr).
\end{equation}

\begin{figure}\label{capacity}
  \begin{center}
  \includegraphics[width=22pc,height=15pc]{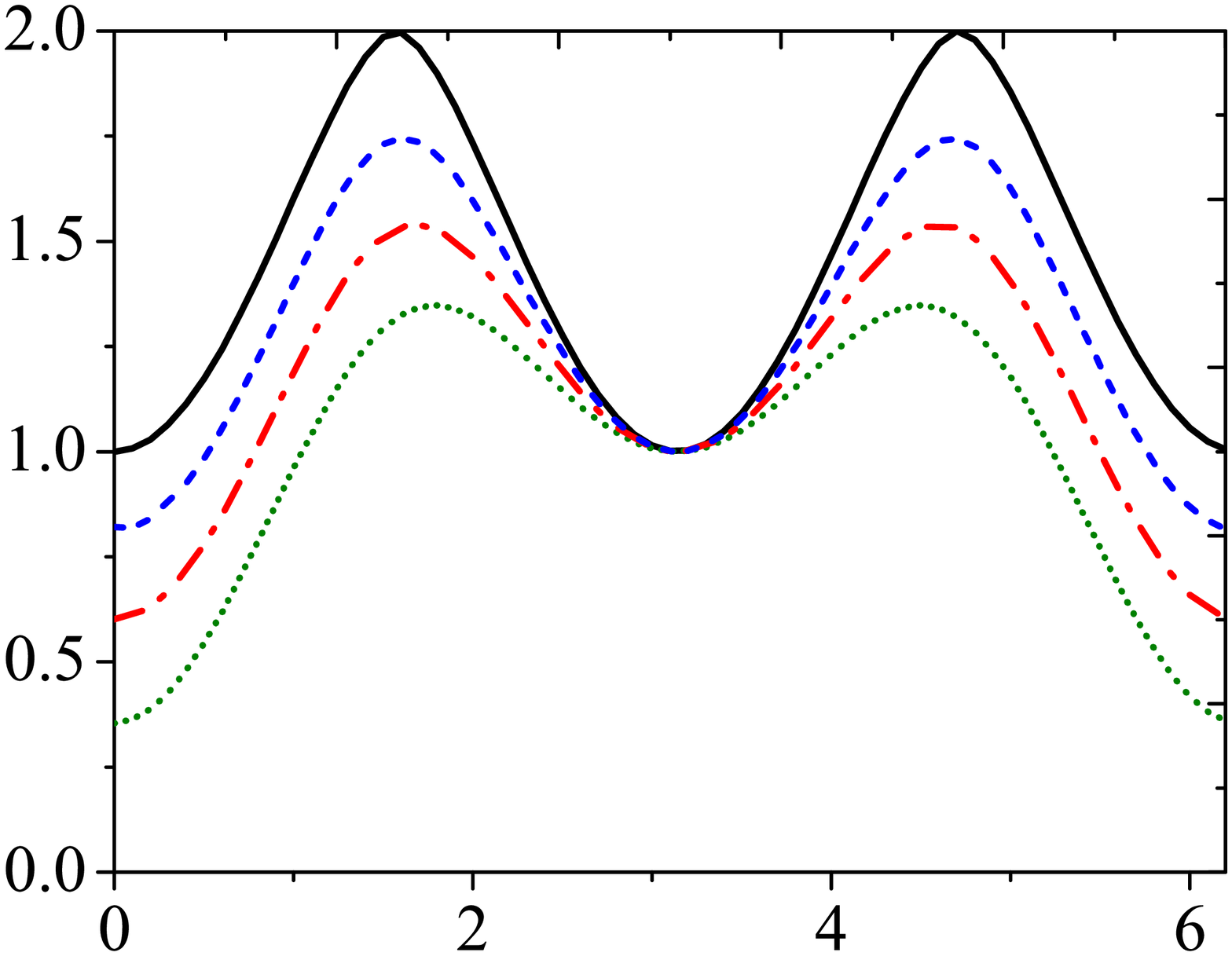}
  \put(-270,90){$\bar{\mathcal{C_P}}^{G}$}
  \put(-130,5){$\Omega'$}
  \includegraphics[width=22pc,height=15pc]{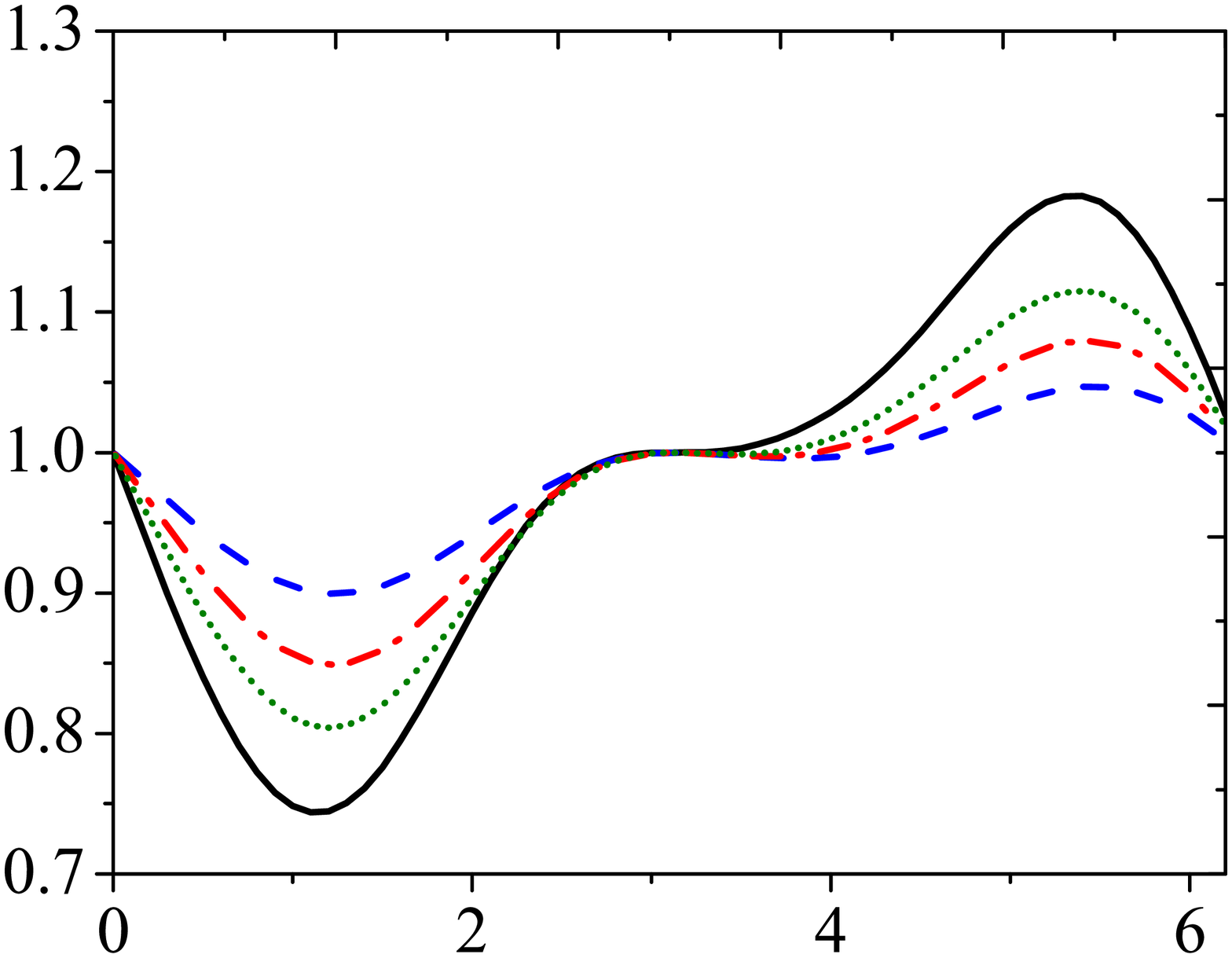}
   \put(-260,90){$\bar{\mathcal{C_P}}^{w}$}
  \put(-130,5){$\Omega'$}
     \caption{The average capacity under Lorentz transformation
     with  $\Omega_1=\Omega_2=\Omega $  (a) for GHZ state (b) for
     W-state.
      The solid, dot, dash-dot and dash  curves for
     and $\Omega_3=0, \frac{\pi}{3},\frac{\pi}{4}$ and $\frac{\pi}{6}$ respectively.}
       \end{center}
\end{figure}

The behavior of the average capacity  $\bar{\mathcal{C_P}}^{g}$ of
the GHZ  state under Lorentz transformation is described in
Fig.(3a) for different values of Wigner's angles. It is assumed
that, the first and second particles are transformed with the same
Wigner angle i.e., $\Omega_1=\Omega_2=\Omega' \in[0,~2\pi]$, while
the third particle is transformed with different Wigner's angle,
$\Omega_3$. The initial average capacity
$\bar{\mathcal{C_P}}^{g}(0,0,\Omega_3)$ of the GHZ state depends
on the stetting value of $\Omega_3$. It is clear that, for small
values of  $\Omega_3$, the initial capacity is large, while it is
small for smaller values of $\Omega_3$. On the other hand, as
$\Omega'$ increases,  the average channel capacity oscillates
between its maximum and minimum bounds. The maximum bounds depend
on the Wigner's angle of the third particle, where for larger
values of $\Omega_3$ the upper bounds of the  average capacity
decrease. However, the minimum bounds are reached at
$\Omega'=\pi$. If we set $\Omega_3=0$, i.e., the third particle is
considered as an observer, the average capacity increases to reach
its maximum bound,i.e., $\bar{\mathcal{C_P}}^{g}=2$ for the first
time at $\Omega'=\frac{5\pi}{12}$.

For W-state the  average channel capacity
$\bar{\mathcal{C_P}}^{w}$ is shown in Fig.$(3b)$, where  the same
 values of the Wigner's angles are used. In this case, the behavior is
completely different. In the interval $[0, \pi]$, the average
capacity decreases as the Wigner's angle of the third particle
$\Omega_3$ decreases. However the lower bounds of
$\bar{\mathcal{C_P}}^{w}$ decrease and consequently the average
capacity increases as $\Omega_3$ increases. This shows that as the
difference between the Wigner's angle is small the average channel
capacity is large. However, in the interval $[\pi,2\pi]$, the
situation is different i.e., the channel capacity increases for
smaller values of $\Omega_3$

From  Fig.(3), one concludes that the average capacity of the GHZ
state may increase if we allow for  one of these particles to play
the role of observer or minimize the  value of Wigner's angle. In
this case the GHZ state can be used to send  a large amount of
information within the context of relativistic quantum
information.

\section{Entanglement} In this section  the effect
of Lorentz transformation on the degree of entanglement of GHZ and
W-states is investigated. For this task we use a measure called
tangle \cite{Minter,Wootter,Dur}. This measure is used to quantify the
three way entanglement.  For a three qubit state
$\ket{\psi}_{123}$, the  three- tangle  of this state is given by,
\begin{equation}
\mathcal{T}_{123}=\mathcal{C}^2_{1(23)}-\mathcal{C}^2_{12}-\mathcal{C}^2_{13},
\end{equation}
where $\mathcal{C}_{1(23)}=2\sqrt{det\{\rho_1\}},\quad
\rho_1=tr_{23}\{\ket{\psi}_{123}\bra{\psi}\}$ and
$\mathcal{C}_{ij}, ij=12,13$ is the concurrence of the two qubit
states $\rho_{12}=tr_{3}\{\ket{\psi}_{123}\bra{\psi}\}$ and
$\rho_{13}=tr_{2}\{\ket{\psi}_{123}\bra{\psi}\}$, respectively.
The concurrence of a two qubit  state $\rho_{ij}$ is defined as
$\mathcal{C}(\rho_{ij})=max\bigl\{0,\sqrt{\lambda_1}-\sqrt{\lambda_2}-\sqrt{\lambda_3}-\sqrt{\lambda_4}~\bigr\}$,
where $\lambda_k, k=1..4$ are the eigenvalues of the matrix
$\rho_{ij}(\sigma_y^{(i)}\otimes\sigma_y^{(j)})\rho_{ij}^*(\sigma_y^{(i)}\otimes\sigma_y^{(j)})$
\cite{Wootter89}.

The effect of the Lorentz transformation  on the degree of
entanglement of the GHZ state is displayed in Fig. $4a$, where it
is assumed that $\Omega_1=\Omega_2=\Omega' \in[0,2\pi]$ and
different values of $\Omega_3$ are considered. It is clear that,
before switching on the Lorentz transformation the tangle
$\mathcal{T}_G=1$. However, as the Wigner angles increase,
entanglement  reach its lower bound at $\Omega'=\pi/3$. For larger
values of $\Omega'$, the tangle $\mathcal{T}_G$ increases again to
reach its maximum value at $\Omega'=4\pi/3$. As $\Omega'$
increases further the entanglement decreases gradually to reach
its lower bound  at $\Omega' \simeq \frac{5\pi}{3}$. Finally the
tangle increases to its maximum value i.e., $\mathcal{T}_G=1$ at
 $\Omega'=2\pi$.
For larger values of $\Omega_3$, the tangle decreases gradually
and the minimum bounds decrease (tangle increase) as the Wigner's
angle $\Omega_3$ increases.

\begin{figure}[b!]\label{EnGHZ}
  \begin{center}
  \includegraphics[width=21pc,height=15pc]{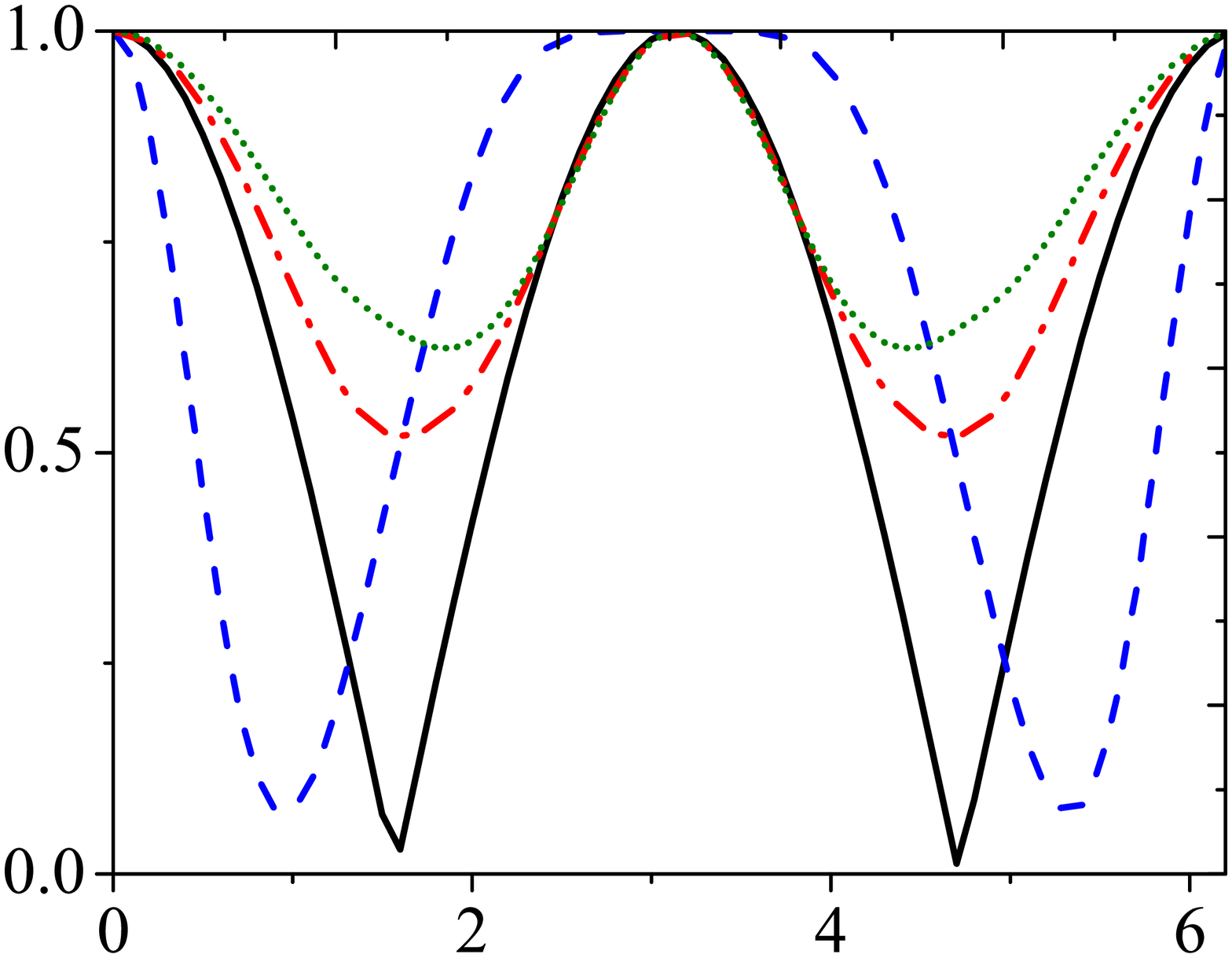}
     \put(-240,90){$\mathcal{T}_G$}
\put(-120,5){$\Omega'$}
 \put(-67,145){$(a)$}
 \includegraphics[width=21pc,height=15pc]{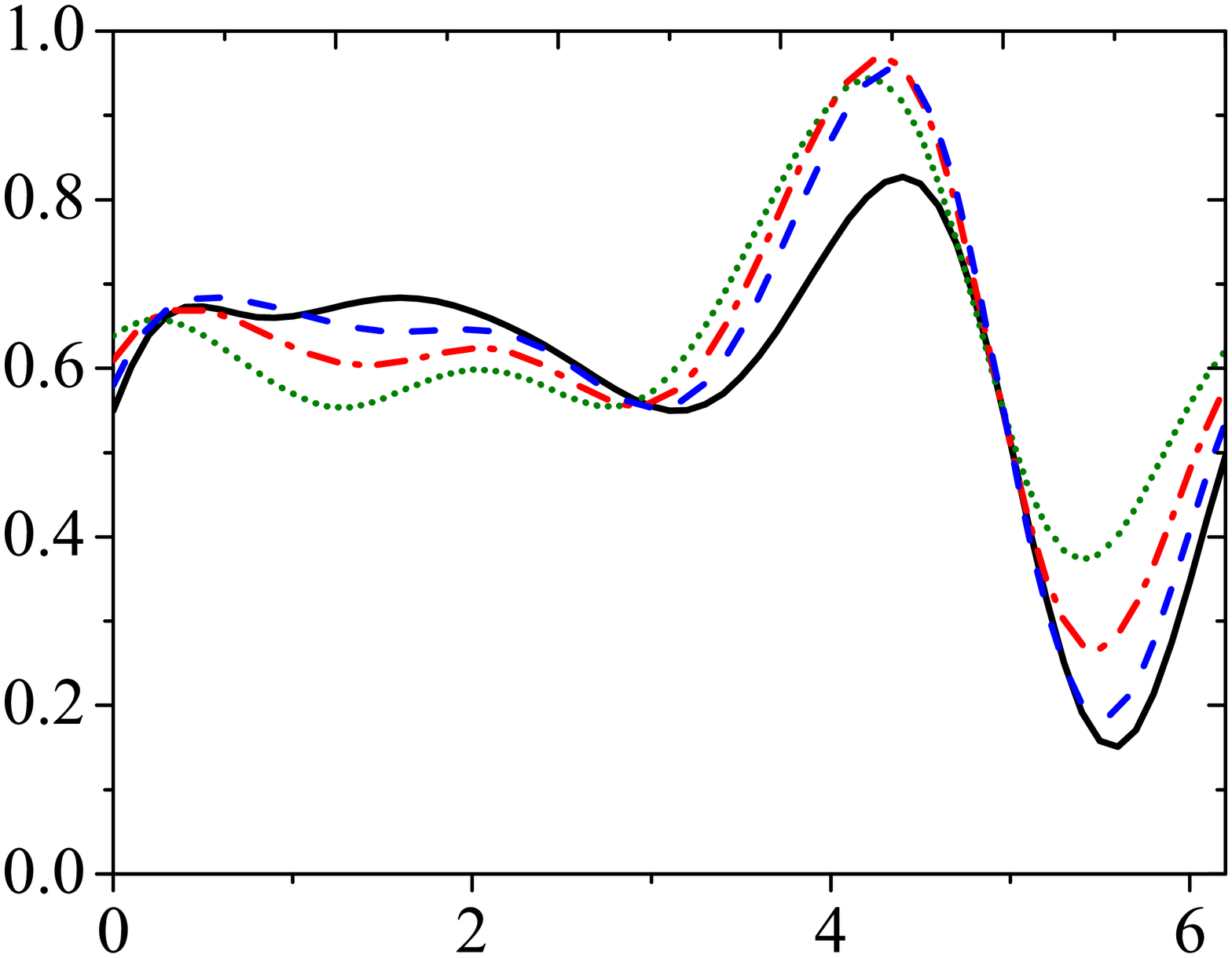}
  \put(-240,90){$\mathcal{T}_{w}$}
  \put(-120,5){$\Omega'$}
 \put(-65,145){$(b)$}
     \caption{The entanglement of(a)  GHZ- state (b) W-state .  The solid, dot, dash-dot and dash  curves for
     and $\Omega_3=0, \frac{\pi}{3},\frac{\pi}{4}$ and $\frac{\pi}{6}$ respectively.}
       \end{center}
\end{figure}

The behavior of the tangle for a system initially prepared in
W-state evolves under the effect of Lorentz transformation is
displayed in Fig.(4b). Jung et.al  \cite{Jung}have shown that the
tangle of W-state, $\mathcal{T}_w \simeq 0.55$ . The general
behavior is similar to that depicted for $\mathcal{T}_G$ as shown
in Fig.(4a). However, for $\Omega'\in[0, \pi]$, the tangle
$\mathcal{T}_w$ increases as the Wigner's angle of the third
particle $\Omega_3$ decreases, the lower bounds are larger than
that shown for $\mathcal{T}_G$. This behavior changes in the
interval $[\pi, 2\pi]$, where for larger values of $\Omega_3$, the
upper bounds of $\mathcal{T}_w$  are larger.

From Fig.4, one concludes that, as the difference between the
Wigner's angle, $\Delta=\Omega'-\Omega_3$ decreases, the tangle of
the GHZ state increases. Due to the structure of W-state, the
tangle increases as the difference between the Wigners's angles
increases in the interval $[0, \pi]$ and decreases as the
difference $\Delta$ increases in the interval $[\pi, 2\pi]$. The
  tangle   behavior shows that the  W-state is more robust than the  GHZ state.

\section{Conclusion}

The effect of the Lorentz transformation on the fidelity, capacity
and entanglement of tripartite systems of GHZ and W-states are
investigated, where the  final  state vectors of  GHZ  and W-
states are obtained analytical as  functions of Wigner angles. The
behavior of these phenomena under the Lorentz transformation, is
considered as a measure of robustness of these states to Lorentz
transformation. It is shown that,
 the  values of these quantities oscillate between
their lower and upper bounds depending on the values of the
Wigner's angles.

The behavior of the fidelities of GHZ and W-states shows that,
these states turn into an equivalent form, where as soon as the
fidelities of the initial state decreases, the fidelity of the
equivalent state re-birthes. However, when the fidelity of the
initial state vanishes completely, the fidelity of the equivalent
state becomes maximum. Therefore, Lorentz transformation keeps the
entangled properties of the transformed states.

The effect of the Lorentz transformation on the channel capacities
shows  different behaviors for GHZ and W-states, where the average
capacity for GHZ increases as the Wigner angles increases, while
it decreases for W-state. For GHZ state, if one particle is
considered as an observer, the upper bounds increase as the
difference between the Wigner's angles of this particle and the
other two particles increases. However, for W-state, the lower
bounds increase (average capacity increases) as the difference
between Wigner's angle increases,.i.e the observer is transformed
with a small Wigner's angle

The amount of survival entanglement is quantified by  means of the
tangle as a measure of entanglement  between three qubits. Our
results show that the tangle decreases  gradually as the Wigner'
angles increase. For GHZ state, the lower bounds decreases, i.e.,
the entanglement increase when the difference between the
observer's Wigner angles and the Wigner angles of the other two
particles decreases, while  for W-states, the entanglement
increases as this difference increases. However, the lower bounds
of entanglement for W-state is much larger than that depicted for
GHZ state. This shows that  the W-state is more robust than GHZ
state under the effect of Lorentz transformation.

 {\it In conclusion:} the entanglement  of the tripartite states can be improved or   almost  kept   invariant,
   if all the particles are transformed with an equal Wigner's
 angles. Since for any Wigner's angles, the fidelity doesn't
 vanish, then these states are transformed to an equivalent form
  at some specific values of Wigner's angle and
 consequently can be used to perform quantum information tasks, as
 teleportation and quantum coding with high efficiency.

\end{document}